\documentstyle[twocolumn,prl,aps,epsf]{revtex}
\begin{document}

\title{Magnon splitting induced by charge ordering in NaV$_2$O$_5$}

\author{Claudius Gros and Roser Valent\'\i} 
\address{   Institut f\"ur
 Physik, Universit\"at Dortmund, 44221 Dortmund, Germany.}
\date{\today}

\maketitle

\begin{abstract}
 We consider the effects of charge ordering 
in NaV$_2$O$_5$ (below $T_{SP}$)
on the exchange constants and on the magnon
dispersion. We show that the experimentally
observed splitting of the magnon branches along 
the $a$ direction
is induced by charge ordering. We find that 
one can distinguish between the proposed 'zig-zag' 
and 'in-line' patterns of charge ordering. 
Only the zig-zag ordering
is consistent with the experimental results regarding
(i) the unusual intensity modulation observed in magnetic 
    neutron scattering,
(ii) the reduction in the
     intra-ladder exchange constant below $T_{SP}$, and
(iii) the magnon dispersion along $a$.
We estimate the inter-ladder exchange constant
to be $1.01\,\mbox{meV}=11.7\,\mbox{K}$
for $T>T_{SP}$.  

\end{abstract}


\centerline{\hfill}

\section*{Introduction}
The physical properties of the quasi one-dimensional (1D)
spin-liquid compound NaV$_2$O$_5$ are currently under
intense investigation. NaV$_2$O$_5$ is an insulator and
its magnetic susceptibility is that of a 1D Heisenberg
chain \cite{Isobe96}. Below $T_{SP}=34\,\mbox{K}$ a
spin-gap opens \cite{Isobe96} and the crystallographic
unit-cell doubles along the chain-direction ($b$ axis)
\cite{Fujii97}.

The nature of the spin-chains was originally supposed
to be an 'in-line' arrangement of $V^{+4}$ spin-1/2 ions
along the crystallographic $b$ direction, 
separated by chains of $V^{+5}$ spin-zero ions
\cite{Isobe96,Galy_Carpy}. Recently though, the notion
of NaV$_2$O$_5$ as a quarter filled ladder compound
has been proposed \cite{Smolinski98}, based on a 
redetermination of the crystal structure
\cite{Smolinski98,Schnering98} and a subsequent
LDA analysis \cite{Smolinski98}. In this interpretation
the electron spins are not localized at V-ions,
but distributed over V-O(1)-V molecules \cite{Smolinski98,Horsch98}, 
the rungs of the constituent V-O ladders running along $b$.
This interpretation
has found support by recent NMR \cite{Ohama98} and
Raman \cite{Popova98} measurements, though far-infrared 
optics have been interpreted as
evidence for charged magnons \cite{Damascelli98}.

The nature of the state below $T_{SP}$ is not
yet determined. Isobe and Ueda \cite{Isobe96}
originally proposed a standard spin-Peierls
scenario. The detection of two inequivalent
V-ions below $T_{SP}$ by NMR \cite{Ohama98}
indicates charge ordering 
in addition to the observed formation of a singlet
ground-state. The pattern of the charge
ordering is currently debated, several
authors have proposed `in-line' and
`zig-zag' patterns, based on electronic
considerations \cite{Seo98,Thalmeier98,Mostovoy98,Riera98}.

The occurence of two well defined magnon-branches
in NaV$_2$O$_5$ along the $a$ direction (perpendicular to the
 chains), as measured by neutron scattering 
\cite{Yosihama98}, has come to a surprise and is
yet unexplained. Here we propose that the
observed splitting of the magnon branch is directly
proportional to the square of the charge order parameter,
in the case of zig-zag ordering. In the case
of in-line charge ordering we find only one magnon branch.
We also observe that zig-zag/in-line charge ordering leads
to a decrease/increase of the magnetic exchange 
below $T_{SP}$ along $b$.
Experimentally a decrease of $\ \approx20\%$ is
observed\cite{Weiden97,Mila96}. We finally comment
on the unusual intensity modulation observed in
neutron-scattering \cite{Yosihama98} and find
it to be consistent/inconsistent with zig-zag/in-line
charge ordering.


\section*{Exchange couplings}

\begin{figure}
   \epsfxsize=0.54\textwidth
\vspace*{-5ex}
\vbox to 7.3cm{\centerline{\epsffile{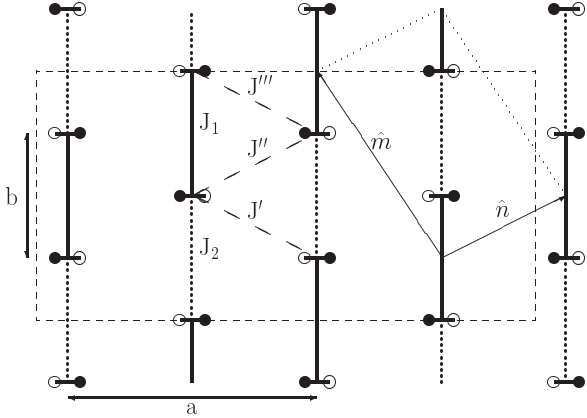}}}

\caption{\label{Fig1}
Illustration of the dimer-phase of NaV$_2$O$_5$
with `zig-zag' charge ordering. V-ions with
charge $(1\pm\Delta_c)/2$ are denoted by
open/filled circles. The spins are in V-O(1)-V 
bonding orbitals (rungs),
denoted by the short horizontal bars connecting
pairs of nearby V-ions
(not on scale, their actual distance is $\sim a/3$).
The dimers are denoted by the vertical bars connecting the
rungs. The lattice constants are $a$ and
$b$. The primitive unit-cell is spanned
by $\hat n$ and $\hat m$ and contains one dimer
($\equiv$ four V-ions). 
The orthorhombic unit-cell (dashed box) contains four dimers.
Note that the in-line charge ordering would correspond
to having parallel chains of V-ions with charge 
$(1\pm\Delta_c)/2$ respectively running along $b$.
        }
\end{figure}

The exchange couplings in the dimerized state ($T<T_{SP}$)
might differ from those in the homogeneous state
($T>T_{SP}$), due to the charge ordering (see Fig.\ \ref{Fig1}).
The sign and the magnitude of these changes depend on
the pattern of the charge ordering. We consider first
the `zig-zag' charge ordering illustrated in Fig.\ \ref{Fig1}. 
We define $\Delta_c$ to be the charge order parameter. 
$(1\pm\Delta_c)/2$ are the charge occupation factors of the two
inequivalent V-ions. $\Delta_c=1$ would correspond to complete
charge disproportionation, $2V^{+4.5}\rightarrow V^{+5}+V^{+4}$.

Following the standard derivation of
the exchange constant $J_{i,j}$
between two V-sites $i$ and $j$ by
perturbation theory, one finds immediately that $J_{i,j}$ is
linearly dependenent on the respective site occupation factors
$\rho_{i/j} = (1\pm\Delta_c)/2$, i.e.\
$J_{i,j}\sim\rho_i\rho_j$. The dominant magnetic
coupling in NaV$_2$O$_5$ is between the rungs
of the constituent ladders along $b$ 
and has contributions from bond-exchange processes
$\sim(\rho_i\rho_j+\rho_k\rho_l)$ and from ring
exchange $\sim\sqrt{\rho_i\rho_j\rho_k\rho_l}$, where
$i,j,k,l$ denote the four sites of the two rungs
\cite{note_1}.
We denote with $J$ and $J_\infty$ the exchange 
constants between rungs along $b$ in
the low- and high-temperature phase respectively. 
The zig-zag charge ordering (see Fig.\ \ref{Fig1}) 
modifies the exchange $J_\infty$ as
$J=J_\infty(1-\Delta_c^2)$
\cite{Smolinski98}. 
Note that the experimental values
for $J\approx 441\,\mbox{K}$ ($T<T_{SP}$)
\cite{Yosihama98,Weiden97} and for
$J_\infty\approx 529-560\,\mbox{K}$ ($T>T_{SP}$) 
\cite{Isobe96,Mila96} indicate
$\Delta_c^2\approx0.2$, i.e.\ $\Delta\approx0.44$ \cite{note_J}.

Another possibility of charge ordering has been
discussed in the literature, where all
charges align along $b$, leading to alternating
$V^{+4}$ and $V^{+5}$ chains \cite{Galy_Carpy}.
For this `in-line' charge ordering the exchange
coupling between rungs along $b$ has contributions
$\sim(1+\Delta_c^2)$ from the bond-exchange processes
and $\sim(1-\Delta_c^2)$ from the ring-exchange. No substantial
reduction of the exchange with charge ordering
is then expected \cite{note_1}.

In Fig.\ \ref{Fig1} we illustrate a dimerization
pattern for NaV$_2$O$_5$ which is consistent with
the observed doubling of the unit-cell along the
$a$ and the $b$ direction \cite{Fujii97}, leading to an
alternation of the exchange coupling $J$ between the rungs
along $b$: $J_{1/2}=J(1\pm\delta)$. 
The modified exchange integrals between rungs
along $a$ due to
the zig-zag charge ordering for $T<T_{SP}$
are (see Fig.\ \ref{Fig1}):
$J^{\prime}=J_\infty^\prime(1+\Delta_c)^2$,
$J^{\prime\prime}=J_\infty^\prime(1-\Delta_c^2)$ and
$J^{\prime\prime\prime}=J_\infty^\prime(1-\Delta_c)^2$,
where $J_\infty^\prime$ is the exchange along $a$
in the high-temperature phase.

\section*{Magnon dispersion}
We now calculate the magnon dispersion using the 
bond-operator theory \cite{Brenig97,Ueda}. As the bond-operator
expansion is a well established technique we refer to the
literature for details and present here only the
basic notions.
The two-electron state on a dimer 
(involving the electrons on rung 1 and rung 2) 
can be in a spin-singlet or in a spin-triplet state
($s^\dagger|0_{(1,2)}\rangle$ and
 $t_\alpha^\dagger|0_{(1,2)}\rangle$ ($\alpha=x,y,z$) 
respectively), e.g.\
\begin{eqnarray*}
s^\dagger|0_{(1,2)}\rangle&=&{1\over\sqrt 2} \left(
|1\uparrow,2\downarrow\rangle -|1\downarrow,2\uparrow\rangle 
                                    \right)~,\\
t_z^\dagger|0_{(1,2)}\rangle&=&{1\over\sqrt 2} \left(
|1\uparrow,2\downarrow\rangle +|1\downarrow,2\uparrow\rangle 
                                    \right)~.
\end{eqnarray*}
%
%

Due to spin-rotational invariance one needs to consider only
the z-component of the inter-dimer
exchange ${\bf S}_l\cdot{\bf S}_{l^\prime}$ 
($l$ and $l^\prime$ are rung-indices)
which couples $s^\dagger|0_{(1,2)}\rangle$ with
$t_z^\dagger|0_{(1,2)}\rangle$ and vice versa.
The intra-dimer matrix elements are
\begin{eqnarray}
\langle 0_{(1,2)}|t_z^{} S_1^z s^\dagger|0_{(1,2)}\rangle&=&+{1\over 2},
\nonumber\\
\langle 0_{(1,2)}|t_z^{} S_2^z s^\dagger|0_{(1,2)}\rangle&=&-{1\over 2}~.
\label{ME}
\end{eqnarray}
The inter-dimer matrix elements of $S_l^zS_{l^\prime}^z$
defining the Hamiltonian in the bond-operator
representation are products of the intra-dimer matrix
elements given in Eq.\ (\ref{ME}) \cite{Brenig97}.

The dimer-covering presented in Fig.\ \ref{Fig1}
contains four dimers in the orthorhombic unit-cell.
It is easier to calculate the dispersion relation
using the primitive unit-cell, which contains only
one dimer. The primitive unit-cell is spanned by
\begin{equation}
\hat n = \left({a\over2},{b\over2}\right),\qquad 
\hat m = \left({-a\over2},{3b\over2}\right)~,
\label{primitive}
\end{equation}
where $a=11.316~\mbox{\AA}$ and $b=3.611~\mbox{\AA}$
are the lattice parameters of the room-temperature
structure\cite{Galy_Carpy,Smolinski98}. The
basis vectors $\hat n$ and $\hat m$ span a simple
square lattice of dimers with couplings
between the dimers in real space
$\tilde J^\prime=2J^{\prime\prime}-J^{\prime\prime\prime}$
along $\pm\hat n$ and
$-J^\prime$ along $\pm\hat m$. 
 In reciprocal space these couplings lead to
momentum dependences
$\sim\cos({\bf k}\cdot\hat n)$ and
$\sim\cos({\bf k}\cdot\hat m)$.
We define the dimensionless matrix element
\begin{eqnarray}
\epsilon_{\bf k} = 
&&{\tilde J^\prime\over2J_1}\cos\left(k_x{a\over2}+k_y{b\over2}\right)
\label{k_vector}\\
&&\qquad-{J^\prime\over2J_1}\cos\left(k_x{a\over2}-k_y{3b\over2}\right)
-{J_2\over2J_1}\cos\left(2bk_y\right)~,
\nonumber
\end{eqnarray}
with ${\bf k}=(k_x,k_y)$. We employ the 
``linearized Holstein-Primakoff'' approximation (LHP)
and eliminate the singlet operators via the constraint that
each dimer can only be in a singlet or in a triplet state.
This procedure leads to the following linearized Hamiltonian 
for the triplet excitations\cite{Brenig97}:
\begin{equation}
J_1\sum_{{\bf k},\alpha}\left[
\left(1+\epsilon_{\bf k}\right) 
t_{{\bf k},\alpha}^\dagger t_{{\bf k},\alpha}^{}
+{\epsilon_{\bf k}\over2} \left(
t_{{\bf k},\alpha}^\dagger t_{-{\bf k},\alpha}^\dagger 
+ 
t_{{\bf k},\alpha}^{} t_{-{\bf k},\alpha}^{} \right) \right]
\label{H}
\end{equation}
The $t_{{\bf k},\alpha}$ obey Bose commutation relations,
$[t_{{\bf k},\alpha}^{},
t_{{\bf k}^\prime,\alpha^\prime}^\dagger]
=\delta_{{\bf k},{\bf k}^\prime}
\delta_{\alpha,\alpha^\prime}$ 
and they are the Fourier
 transform of the triplet Bose operators on the dimer bonds.
Eq.\ (\ref{H}) is solved by a standard Bogoliubov
transformation \cite{Brenig97}, 
leading to the (threefold degenerate)
magnon-dispersion
\begin{equation}
E({\bf k}) = J_1\sqrt{1+2\epsilon_{\bf k}}
\label{E_k}~.
\end{equation}
Yosihama {\it et al.} \cite{Yosihama98} have measured 
$E({\bf k})$ by inelastic neutron scattering
and found a splitting for $k_y=\pi/b$.
Specializing Eq.\ ({\ref{E_k}) to $k_y=\pi/b$ we obtain
\begin{equation}
E(k_x,{\pi\over b}) = \sqrt{J_1(J_1-J_2)+
J_1(\tilde J^\prime-J^\prime)\cos({k_xa+\pi\over2})
                     }
\label{E_exp}~.
\end{equation}
$E(k_x,\pi/b)$ has the periodicity $4\pi/a$. Folding
back to the first Brillouin zone 
($k_x\in[-\pi/a,\pi/a]$)
we have two magnon
branches which are degenerate at $k_x=0$ 
and split at $k_x=\pi/a$,
as seen in experiment \cite{Yosihama98,note_2}.
Note that the dispersion along $a$ (and the splitting)
is possible only for a finite $\Delta_c\ne0$, since
\begin{equation}
\tilde J^\prime-J^\prime =
2J^{\prime\prime}-J^{\prime\prime\prime}-J^\prime 
= J_\infty^\prime(-4\Delta_c^2)
\approx -0.8J_\infty^\prime~,
\label{dis}
\end{equation}
where we have used $\Delta_c^2\approx0.2$. 
The fact that the magnon
dispersion along $a$ is $\sim\Delta_c^2$ for $k_y=0,\pi/b$
(in leading order in $J_\infty^\prime$) is a direct 
consequence of the complete mutual frustration of 
n.n.\ ladders in NaV$_2$O$_5$.

   \begin{figure}[tbh]
   \epsfxsize=0.48\textwidth
   \centerline{\epsffile{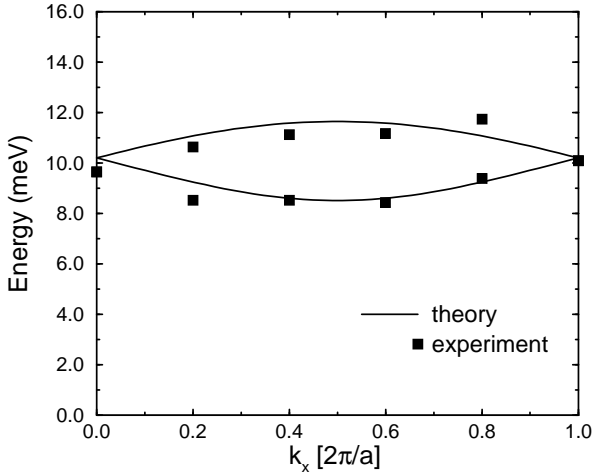}}
   \medskip
   \caption{\label{Fig2}
   Theory (solid line) and experimental (filled squares)
   results \protect\cite{Yosihama98}
   for the magnon dispersion in NaV$_2$O$_5$.
           }
   \end{figure}

The experimentally observed gap of 
$\Delta E=10\,\mbox{meV}$
at $k_x=0$ determines $\delta$:
\begin{equation}
\Delta E=J\sqrt{2(1+\delta)\delta},\qquad
\delta \sim 0.034~,
\label{delta}
\end{equation}
where we have used $J=440\,\mbox{K}=37.9\,\mbox{meV}$.
This value of $\delta$
should be considered as a lower bound to the true
value of $\delta$, since the LHP overestimates somewhat
the gap for small values of the exchange alternation
$\delta$ \cite{Brenig97}. The splitting 
$E_2-E_1=3.2\,\mbox{meV}$ \cite{Yosihama98} 
of the magnon branch at $k_x=\pi/a$ is given by
\begin{equation}
E_2-E_1=E(\pi/a,\pi/b)-E(3\pi/a,\pi/b)~.
\label{splitting}
\end{equation}
Using Eq.\ (\ref{E_exp})  and
taking $\Delta_c^2=0.2$ we find
$J_\infty^\prime=1.01\,\mbox{meV}
                =11.7\,\mbox{K}$.
The anisotropy
is therefore roughly $J_\infty^\prime/J_\infty\sim 1/45$.
This small value of the inter-ladder coupling is consistent
with an argument by Horsch {\it et al.}
\cite{Horsch98}, that 
there is a partial cancellation between intermediate
singlet and triplet states in
the superexchange paths contributing to $J_\infty^\prime$.
In Fig.\ \ref{Fig2} we compare the resulting magnon dispersion
with the experimental result. 

\begin{figure}[bth]
\epsfxsize=0.54\textwidth
\vspace*{-12ex}
\vbox to 8.0cm{\centerline{\epsffile{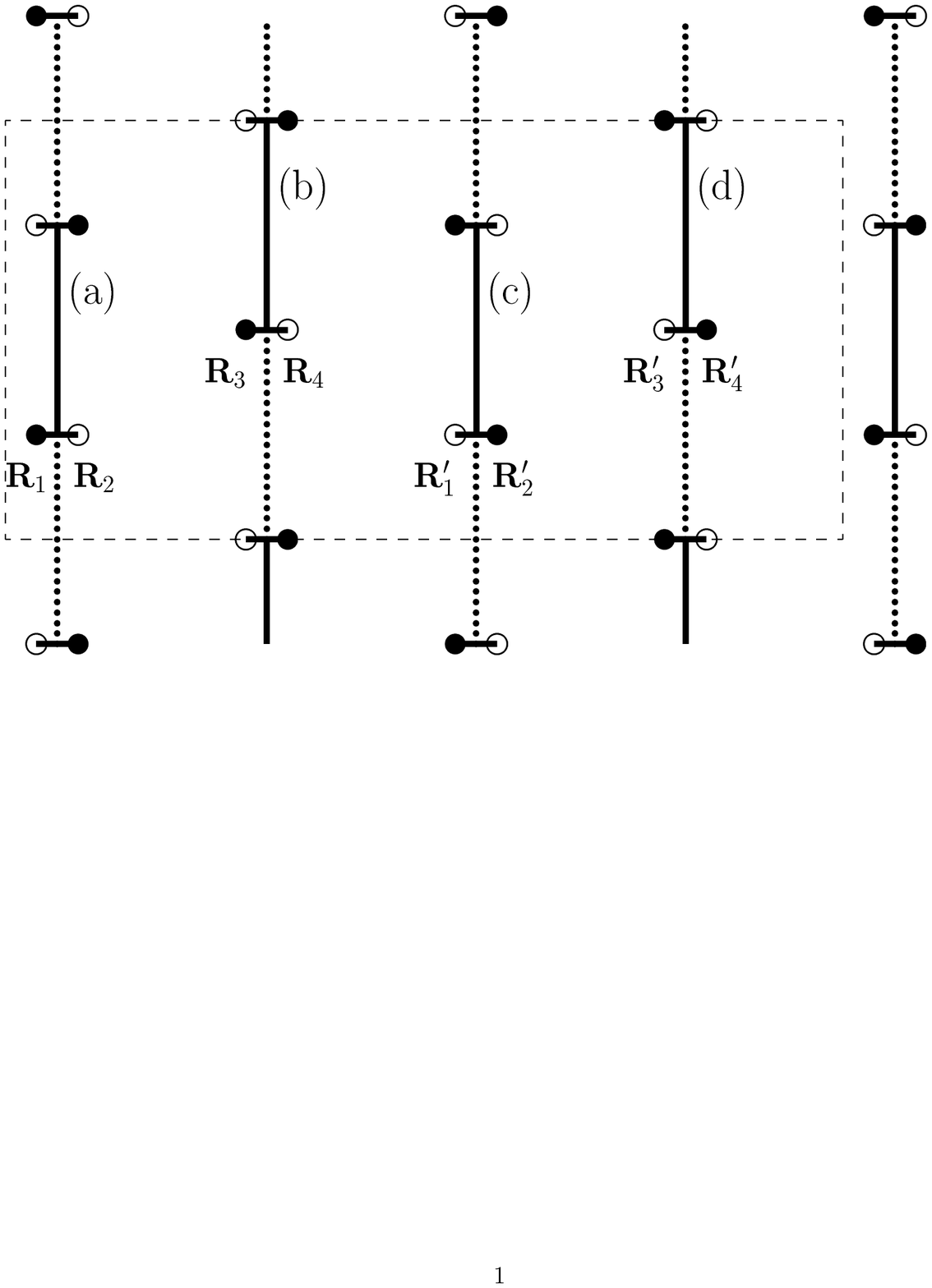}}}

\caption{\label{Fig3}
A second possible dimer configuration for the dimerized
phase of NaV$_2$O$_5$. The symbols are the same as in
Fig.\ 1. Due to the charge ordering the dimers
denoted with (a) and (b) are not equivalent 
to the dimers denoted with (c) and (d).
The ${\bf R}_l$ and ${\bf R}_l^\prime$ denote the
positions of the V-ions (not on scale).
  }
\end{figure}

\section*{discussion}
We have seen that the observed splitting of the
magnon branches along the $a$ direction 
in NaV$_2$O$_5$ can only be explained
if the total coupling between n.n.\ ladders,
$J_{total}^\prime=\tilde J^\prime-J^\prime = 2J^{\prime\prime}-
J^{\prime\prime\prime}-J^\prime$ does not sum up
to zero. As a possible source on nonzero
$J_{total}^\prime$ we have identified the charge ordering
below $T_{SP}$ \cite{Ohama98}. 
Another possible candidate for a
nonzero $J_{total}^\prime$ could be, in principle,
the structural distortions occuring for
$T<T_{SP}$. There are two types of structural
distortions: (i) distortions $\sim\delta$ leading
to the dimerization along $b$ and (ii) distortions
leading to two inequivalent V-ions. This last type of
distortions, which are the underlying cause for the charge 
disproportionation occuring for $T<T_{SP}$, could  
also alter the exchange 
constant directly, contributing to a 
nonzero $J_{total}^\prime$.
At this point we cannot rule out this possibility, which
however would not affect the results presented.
Let us also remark in this context, that the originally
proposed in-line charge ordering \cite{Galy_Carpy,Isobe96}
would result in 
$J^\prime=J^{\prime\prime}=J^{\prime\prime\prime}$
and thus in a vanishing $J_{total}^\prime$. An
in-line charge ordering can therefore not 
explain the existence of two distinct
magnon branches in NaV$_2$O$_5$.

We have seen that the dimerization pattern shown
in Fig.\ \ref{Fig1} is able to explain the measured
magnon dispersion in detail. We note here,
that another pattern (illustrated in Fig.\ \ref{Fig3})
is also able to explain the experimental data. The
calculations are, however, more complicated since there
are four dimers in the unit-cell, complicating the
Bogoliubov transformation. We have solved the resulting
$8\times8$ matrix numerically and found for $k_y=\pi/b$ 
results identical to those presented
in Fig.\ \ref{Fig2}. For a general value of $k_y$
the magnon dispersion arising from the two
dimer patterns differ slightly. 
Besides the two dimer coverings illustrated in
Fig.\ (\ref{Fig1}) and Fig.\ (\ref{Fig3}) dimer
coverings involving inter-ladder dimers are
in principle possible. We do not consider them
here, since the inter-ladder exchange $J_\infty^\prime$ 
is very small.

Next we discuss the interpretation of the
unusual modulation of the intensity of the magnon branch
found in neutron-scattering \cite{Yosihama98}.
The authors of Ref.\ \cite{Yosihama98} found that
this intensity has a period of $3\pi/a$ in
$k_x$. The authors argued that this intensity
modulation could be understood only if there is
substantial magnetic scattering from both legs 
of the structural ladders\cite{note_3}, in
accordance with the notion that NaV$_2$O$_5$ 
is a quarter-filled ladder compound
\cite{Smolinski98}. 

The intensity modulation $I_{mod.}$
is proportional to the square of the
weighted sum of the scattering
phases over the magnetic ions in the 
orthorhombic unit-cell.
We denote the positions of the V-ions by
${\bf R}_1,\dots{\bf R}_4$ and 
${\bf R}_1^\prime,\dots{\bf R}_4^\prime$
(see Fig.\ \ref{Fig3}). 
\begin{equation}
I_{mod.}\sim \bigg|
\sum_{l=1}^4\,\rho_l\,\mbox{e}^{i{\bf k}\cdot{\bf R}_l}
+\sum_{l=1}^4\,\rho_l^\prime\,\mbox{e}^{i{\bf k}\cdot{\bf R}_l^\prime}
\bigg|^2~.
\label{intensity}
\end{equation}
The ${\bf R}_l$ and ${\bf R}_l^\prime$ are equivalent,
i.e.\ 
$\exp[i{\bf k}\cdot{\bf R}_l]
=\exp[i{\bf k}\cdot{\bf R}_l^\prime]$. For the 
zig-zag ordering the charge occupation factors
$\rho_l=(1\pm\Delta_c)/2$\ obey
$\rho_l+\rho_l^\prime\equiv1$. The charge order
parameter $\Delta_c$ does therefore drop out
of Eq.\ (\ref{intensity}) and the above
expression for $I_{mod.}$ reduces to the one
proposed in Ref.\ \cite{Yosihama98}. The 
zig-zag ordering is therefore in agreement with
experiment. $I_{mod.}$ would be substantially reduced
for the in-line charge ordering, which has
$\rho_l=\rho_l^\prime$. The in-line charge
ordering would therefore not be in accordance
with the neutron-scattering results presented
in Ref.\ \cite{Yosihama98}.

\section*{Conclusions}
We have discussed two possible types of charge
ordering and found that both the unusual
intensity modulation found in magnetic neutron
scattering \cite{Yosihama98} and the reduction
of the exchange constant along $b$ below $T_{SP}$
indicate `zig-zag' charge ordering. We have examined
the effects of charge ordering upon the magnon dispersion
along $a$ and found, with a bond-operator study,
that the splitting of the magnon branches 
found in experiment \cite{Yosihama98} is a direct
consequence of the charge ordering below $T_{SP}$.
The theory presented here consequently 
predicts the splitting
$E_2-E_1$ to vanish as $T\rightarrow T_{SP}$
(from below), as the charge-order vanishes
for $T>T_{SP}$ \cite{Smolinski98,Schnering98,Ohama98}

We would like to note that we predict a phase shift
of $\pi/2$ between the 
two magnon branches for $k_y=\pi/b$
and $k_y=0, 2\pi/b$ (see Eq.\ (\ref{k_vector})).
For $k_y=\pi/b$ the magnons are degenerate at
$k_x=0$ and split at $k_x=\pi/a$. For $k_y=0,2\pi/b$
the magnons split at $k_x=0$ and are degenerate 
at $k_x = \pi/a$.

We would like to acknowledge discussions with 
P. Lemmens and the support of the German Science
Foundation.



\begin{thebibliography}{99}
\bibitem{note_1} The bond-exchange is $\sim t_\parallel^2/U$
                 and the ring exchange $\sim t_\parallel^2/E_c$,
                 where $E_c$ is the charge-transfer gap
                 \protect\cite{Smolinski98,Horsch98,Mostovoy98}.
                 Both contributions are roughly comparable for
                 NaV$_2$O$_5$.
\bibitem{note_J} There is still a certain uncertainty in
                 the experimental data for $J$ and $J_\infty$,
                 which, however, do not affect the conclusions
                 of this paper.
\bibitem{note_2} We follow Ref.\ \protect\cite{Yosihama98} and
                 use the high-temperature Brillouin zone.
\end{thebibliography}
\end{document}